\documentclass[12pt]{article}
\pdfoutput=1

\usepackage{amsmath}
\usepackage{amssymb}
\usepackage{epsfig}
\usepackage{epstopdf}
\usepackage{latexsym}
\usepackage{color}
\numberwithin{equation}{section}
\usepackage[
      colorlinks=false,
      linkcolor=darkblue,  
      urlcolor=blue,    
      filecolor=blue,     
      citecolor=red,
linktocpage=true,
      pdfstartview=FitV,
      bookmarksopen=true    
      ]{hyperref}

\usepackage[bbgreekl]{mathbbol}
\DeclareSymbolFontAlphabet{\mathbbl}{bbold}

\begin{document}

\begin{titlepage}

\centerline
\centerline
\centerline
\bigskip
\centerline{\Huge \rm M5-branes and D4-branes wrapped on a direct} 
\bigskip
\centerline{\Huge \rm product of spindle and Riemann surface} 
\bigskip
\bigskip
\bigskip
\bigskip
\bigskip
\bigskip
\bigskip
\bigskip
\centerline{\rm Minwoo Suh}
\bigskip
\centerline{\it Department of Physics, Kyungpook National University, Daegu 41566, Korea}
\bigskip
\centerline{\tt minwoosuh1@gmail.com} 
\bigskip
\bigskip
\bigskip
\bigskip
\bigskip
\bigskip
\bigskip
\bigskip
\bigskip
\bigskip
\bigskip

\begin{abstract}
\noindent We construct multi-charged $AdS_3\times\mathbbl{\Sigma}\times\Sigma_{\mathfrak{g}}$ and $AdS_2\times\mathbbl{\Sigma}\times\Sigma_{\mathfrak{g}}$ solutions from M5-branes and D4-branes wrapped on a direct product of spindle, $\mathbbl{\Sigma}$, and Riemann surface, $\Sigma_{\mathfrak{g}}$. Employing uplift formula, we obtain these solutions by uplifting the multi-charged $AdS_3\times\mathbbl{\Sigma}$ and $AdS_2\times\mathbbl{\Sigma}$ solutions to seven and six dimensions, respectively. We further uplift the solutions to eleven-dimensional and massive type IIA supergravity and calculate the holographic central charge and the Bekenstein-Hawking entropy, respectively. We perform the gravitational block calculations and, for the $AdS_3\times\mathbbl{\Sigma}\times\Sigma_{\mathfrak{g}}$ solutions, the result precisely matches the holographic central charge from the supergravity solutions.
\end{abstract}

\vskip 4cm

\flushleft {June, 2022}

\end{titlepage}

\tableofcontents

\section{Introduction}

Recently, there was a discovery of novel class of anti-de Sitter solutions obtained from branes wrapped on an orbifold, namely, a spindle, \cite{Ferrero:2020laf}. The spindle, $\mathbbl{\Sigma}$, is an orbifold, $\mathbbl{WCP}_{[n_-,n_+]}^1$, with conical deficit angles at two poles. The spindle numbers, $n_-$, $n_+$, are arbitrary coprime positive integers. Interestingly, these solutions realize the supersymmetry in different ways from very well studied topological twist in field theory, \cite{Witten:1988ze}, and in gravity, \cite{Maldacena:2000mw}. It was first constructed from D3-branes, \cite{Ferrero:2020laf, Hosseini:2021fge, Boido:2021szx}, and then generalized to other branes: M2-branes, \cite{Ferrero:2020twa, Cassani:2021dwa, Ferrero:2021ovq, Couzens:2021rlk}, M5-branes, \cite{Ferrero:2021wvk}, and D4-branes, \cite{Faedo:2021nub, Giri:2021xta}. Furthermore, two possible ways of realizing supersymmetry, topologically topological twist and anti-twist, were studied, \cite{Ferrero:2021etw, Couzens:2021cpk}.

The spindle solutions were then generalized to an orbifold with a single conical deficit angle, namely,  a topological disk. These solutions were first constructed from M5-branes, \cite{Bah:2021mzw, Bah:2021hei}, and proposed to be a gravity dual to a class of 4d $\mathcal{N}=2$ Argyres-Douglas theories, \cite{Argyres:1995jj}. See also \cite{Couzens:2022yjl} for further generalizations. Brane solutions wrapped on a topological disk were then constructed from D3-branes, \cite{Couzens:2021tnv, Suh:2021ifj}, M2-branes, \cite{Suh:2021hef, Couzens:2021rlk}, D4-branes, \cite{Suh:2021aik}, and more from M5-branes, \cite{Karndumri:2022wpu}. See also \cite{Gutperle:2022pgw} for defect solutions from different completion of global solutions.

An interesting generalization would be to find $AdS$ solutions from branes wrapped on an orbifold of dimensions more than two. Four-dimensional orbifolds are natural place to look for such constructions and some solutions were found. First, by uplifting $AdS_3\times\mathbbl{\Sigma}$ solutions, where $\mathbbl{\Sigma}$ is a spindle, \cite{Boido:2021szx}, or a disk, \cite{Suh:2021ifj}, $AdS_3\times\mathbbl{\Sigma}\times\Sigma_{\mathfrak{g}}\times{S}^4$ solutions from M5-branes were obtained where $\Sigma_{\mathfrak{g}}$ is a Riemann surface of genus $\mathfrak{g}$. Also $AdS_2\times\mathbbl{\Sigma}\times\Sigma_{\mathfrak{g}}$ solutions with spindle, $\mathbbl{\Sigma}$, from D4-branes were obtained, \cite{Giri:2021xta, Faedo:2021nub}. More recently, performing and using a consistent truncation on a spindle, $AdS_3\times\mathbbl{\Sigma}_1\ltimes\mathbbl{\Sigma}_2$ solutions from M5-branes wrapped on a spindle fibered over another spindle were found, \cite{Cheung:2022ilc}. Also $AdS_3\times\mathbbl{\Sigma}\ltimes\Sigma_{\mathfrak{g}}$ solutions on a spindle fibered over Riemann surface were found, \cite{Cheung:2022ilc}.

In this work, we fill in the gaps in the literature. First, we construct multi-charged $AdS_3\times\mathbbl{\Sigma}\times\Sigma_{\mathfrak{g}}$ solutions from M5-branes. Employing the consistent truncation of \cite{MatthewCheung:2019ehr}, we obtain the solutions by uplifting the multi-charged $AdS_3\times\mathbbl{\Sigma}$ solutions, \cite{Boido:2021szx}, to seven-dimensional gauged supergravity. When the solutions are uplifted to eleven-dimensional supergravity, they precisely match the previously known $AdS_3\times\mathbbl{\Sigma}\times\Sigma_{\mathfrak{g}}\times{S}^4$ solutions in \cite{Boido:2021szx} and \cite{Suh:2021ifj}, which were obtained by uplifting the $AdS_3\times\mathbbl{\Sigma}$ solutions of five-dimensional gauged supergravity. However, it is the first time to construct the $AdS_3\times\mathbbl{\Sigma}\times\Sigma_{\mathfrak{g}}$ solutions in seven-dimensional gauged supergravity.

Second, we construct multi-charged $AdS_2\times\mathbbl{\Sigma}\times\Sigma_{\mathfrak{g}}$ solutions from D4-branes. Inspired by the consistent truncation in \cite{Hosseini:2020wag}, we construct them by uplifting the multi-charged $AdS_2\times\mathbbl{\Sigma}$ solutions, \cite{Ferrero:2021etw}, to matter coupled $F(4)$ gauged supergravity. Our solutions generalize the minimal $AdS_2\times\mathbbl{\Sigma}\times\Sigma_{\mathfrak{g}}$ solutions in \cite{Faedo:2021nub} and also the solutions obtained in \cite{Giri:2021xta}. We then uplift the solutions to massive type IIA supergravity to obtain $AdS_2\times\mathbbl{\Sigma}\times\Sigma_{\mathfrak{g}}\times\tilde{S}^4$.

Finally, we perform the gravitational block calculations and, for the $AdS_3\times\mathbbl{\Sigma}\times\Sigma_{\mathfrak{g}}$ solutions, the result precisely matches the holographic central charge obtained from the supergravity solutions.

In section \ref{sec2}, we construct $AdS_3\times\mathbbl{\Sigma}\times\Sigma_{\mathfrak{g}}$ solutions from M5-branes. We uplift the solutions to eleven-dimensional supergravity and calculate the holographic central charge. In section \ref{sec3}, we construct $AdS_2\times\mathbbl{\Sigma}\times\Sigma_{\mathfrak{g}}$ solutions from D4-branes. We uplift the solutions to massive type IIA supergravity and calculate the Bekenstein-Hawking entropy. In section \ref{sec4}, we present the gravitational block calculations. In section \ref{sec5}, we conclude. We present the equations of motion in appendix \ref{appA} and briefly review the consistent truncations of \cite{MatthewCheung:2019ehr} in appendix \ref{appB}.

\section{M5-branes wrapped on $\mathbbl{\Sigma}\times\Sigma_{\mathfrak{g}}$} \label{sec2}

\subsection{$U(1)^2$-gauged supergravity in seven dimensions}

We review $U(1)^2$-gauged supergravity in seven dimensions, \cite{Liu:1999ai}, in the conventions of \cite{Cheung:2022ilc}. The bosonic field content is consist of the metric, two $U(1)$ gauge fields, $A^{12}$, $A^{34}$, a three-form field, $S^5$, and two scalar fields, $\lambda_1$, $\lambda_2$. The Lagrangian is given by
\begin{align} \label{sevenlag}
\mathcal{L}\,=\,&\left(R-V\right)\text{vol}_7-6*d\lambda_1\wedge\,d\lambda_1-6*d\lambda_2\wedge\,d\lambda_2-8*d\lambda_1\wedge\,d\lambda_2 \notag \\
&-\frac{1}{2}e^{-4\lambda_1}*F^{12}\wedge\,F^{12}-\frac{1}{2}e^{-4\lambda_2}*F^{34}\wedge\,F^{34}-\frac{1}{2}e^{-4\lambda_1-4\lambda_2}*S^5\wedge\,S^5 \notag \\
&+\frac{1}{2g}S^5\wedge\,dS^5-\frac{1}{g}S^5\wedge\,F^{12}\wedge\,F^{34}+\frac{1}{2g}A^{12}\wedge\,F^{12}\wedge\,F^{34}\wedge\,F^{34}\,,
\end{align}
where $F^{12}=dA^{12}$, $F^{34}=dA^{34}$ and the scalar potential is
\begin{equation}
V\,=\,g^2\left[\frac{1}{2}e^{-8\left(\lambda_1+\lambda_2\right)}-4e^{2\left(\lambda_1+\lambda_2\right)}-2e^{-2\left(2\lambda_1+\lambda_2\right)}-2e^{-2\left(\lambda_1+2\lambda_2\right)}\right]\,.
\end{equation}
The equations of motion are presented in appendix \ref{appA}.

\subsection{Multi-charged $AdS_3\times\mathbbl{\Sigma}$ solutions} \label{ads3sig}

We review the $AdS_3\times\mathbbl{\Sigma}$ solutions of $U(1)^3$-gauged $\mathcal{N}=2$ supergravity in five dimensions, \cite{Boido:2021szx}. These solution are obtained from D3-branes wrapped on a spindle, $\mathbbl{\Sigma}$. The metric, gauge fields and scalar fields read
\begin{align}
ds_5^2\,=&\,H^{1/3}\left[ds_{AdS_3}^2+\frac{1}{4P}dy^2+\frac{P}{H}dz^2\right]\,, \notag \\
A^{(I)}\,=&\,\frac{y-\alpha}{y+3K_I}dz\,, \qquad X^{(I)}\,=\,\frac{H^{1/3}}{y+3K_I}\,,
\end{align}
where $I=1,\ldots,3$ and the functions are defined to be
\begin{equation}
H\,=\,\left(y+3K_1\right)\left(y+3K_2\right)\left(y+3K_3\right)\,, \qquad P\,=\,H-\left(y-\alpha\right)^2\,,
\end{equation}
where $K_I$ and $\alpha$ are constant and satisfy the constraint, $K_1+K_2+K_3=0$. 

In the case of three distinct roots, $0<y_1<y_2<y_3$, of cubic polynomial, $P(y)$, the solution is positive and regular in $y\in\left[y_1,y_2\right]$. The spindle, $\mathbbl{\Sigma}$, is an orbifold, $\mathbbl{WCP}_{[n_-,n_+]}^1$, with conical deficit angles at $y\,=\,y_1,\,y_2$, \cite{Boido:2021szx}. The spindle numbers, $n_-$, $n_+$, are arbitrary coprime positive integers. The Euler number of the spindle is given by
\begin{equation}
\chi(\mathbbl{\Sigma})\,=\,\frac{1}{4\pi}\int_{\mathbbl{\Sigma}}R_{\mathbbl{\Sigma}}\text{vol}_{\mathbbl{\Sigma}}\,=\,\frac{n_-+n_+}{n_-n_+}\,,
\end{equation}
where $R_{\mathbbl{\Sigma}}$ and $\text{vol}_{\mathbbl{\Sigma}}$ are the Ricci scalar and the volume form on the spindle. The magnetic flux through the spindle is given by
\begin{equation} \label{defp5}
Q_I\,=\,\frac{1}{2\pi}\int_{\mathbbl{\Sigma}}F^{(I)}\,=\,\frac{\left(y_2-y_1\right)\left(\alpha+3K_I\right)}{\left(y_1+3K_I\right)\left(y_2+3K_I\right)}\frac{\Delta{z}}{2\pi}\,\equiv\,\frac{p_I}{n_-n_+}\,,
\end{equation}
and we demand $p_I\in\mathbb{Z}$. One can show that the R-symmetry flux is given by
\begin{equation} \label{rsymm5}
Q_1+Q_2+Q_3\,=\,\frac{p_1+p_2+p_3}{n_-n_+}\,=\,\frac{\eta_1n_+-\eta_2n_-}{n_-n_+}\,,
\end{equation}
where the supersymmetry is realized by, \cite{Ferrero:2021etw},
\begin{align} \label{etaeta}
\text{Anti-twist}: \qquad (\eta_1,\eta_2)\,=\,(+1,+1)\,, \notag \\
\text{Twist}: \qquad \qquad (\eta_1,\eta_2)\,=\,(-1,+1)\,.
\end{align}
In minimal gauged supergravity, $K_1\,=\,K_2\,=\,K_3$, only the anti-twist solutions are allowed. Otherwise, both anti-twist and twist are allowed.

One can express $\Delta{z}$, $y_1$, $y_2$, and the parameters, $K_I$, $\alpha$, in terms of the spindle numbers, $n_-$, $n_+$, $p_1$, and $p_2$, \cite{Boido:2021szx}. The period of the coordinate, $z$, is given by
\begin{equation}
\frac{\Delta{z}}{2\pi}\,=\,\frac{\left(n_--n_+\right)\left(p_1+p_2\right)+n_-n_+-p_1^2-p_1p_2-p_2^2}{n_-n_+\left(n_-+n_-\right)}\,,
\end{equation}
In the special case of
\begin{equation} \label{specialcase}
K_1\,=\,K_2\,, \qquad X^{(1)}\,=\,X^{(2)}\,, \qquad A^{(1)}\,=\,A^{(2)}\,, 
\end{equation}
expressions of $y_1$, $y_2$, and $K_1\,=\,K_2$ are simpler,
\begin{align} \label{ads7y1y2}
y_1\,=&\,\frac{q\left(n_++q\right)\left[2n_-^2-2n_-\left(n_++4q\right)+q\left(5n_++9q\right)\right]}{3\left[n_-\left(n_++2q\right)-q\left(2n_++3q\right)\right]^2}\,, \notag \\
y_2\,=&\,-\frac{q\left(n_--q\right)\left[2n_+^2-2n_+\left(n_--4q\right)-q\left(5n_--9q\right)\right]}{3\left[n_-\left(n_++2q\right)-q\left(2n_++3q\right)\right]^2}\,, \notag \\
K_1\,=&\,K_2\,=\,\frac{q\left(n_--n_+-3q\right)\left(n_++q\right)\left(n_--q\right)}{9\left[n_-\left(n_++2p\right)-q\left(2n_++3q\right)\right]^2}\,,
\end{align}
where we define $q\,\equiv\,p_1\,=\,p_2$. For the expression of $\alpha$, we leave the readers to \cite{Boido:2021szx}. For this special case, the $AdS_3\times\mathbbl{\Sigma}$ solutions are also solutions of $SU(2)\times\,U(1)$-gauged $\mathcal{N}=4$ supergravity in five dimensions, \cite{Romans:1985ps}. The solutions can be uplifted to eleven-dimensional supergravity, \cite{Gauntlett:2007sm}, as it was done for a spindle, \cite{Boido:2021szx}, and for a disk, \cite{Suh:2021ifj}.

\subsection{Multi-charged $AdS_3\times\mathbbl{\Sigma}\times\Sigma_{\mathfrak{g}}$ solutions}

A consistent reduction of seven-dimensional maximal gauged supergravity, \cite{Pernici:1984xx}, on a Riemann surface was performed in \cite{MatthewCheung:2019ehr}. Empolying the consistent truncation, we uplift the $AdS_3\times\mathbbl{\Sigma}$ solutions in section \ref{ads3sig} with 
\begin{equation}
K_1=K_2\ne{K}_3\,,
\end{equation}
to $U(1)^2$-gauged supergravity in seven dimension. We briefly summarize the uplift by consistent truncation in appendix \ref{appB}. As a result, we find the $AdS_3\times\mathbbl{\Sigma}\times\Sigma_{\mathfrak{g}}$ solutions,
\begin{align} \label{ads3sigsigsol}
ds_7^2\,=&\,e^{-4\varphi}H^{1/3}\left(ds_{AdS_3}^2+\frac{1}{4P}dy^2+\frac{P}{H}dz^2\right)+\frac{1}{g^2}e^{6\varphi}ds_{{\Sigma}_{\mathfrak{g}}}\,, \notag \\
e^{-\frac{10}{9}\lambda_1}\,=&\,2^{1/3}X\,, \qquad e^{\frac{5}{3}\lambda_2}\,=\,2^{1/3}X\,, \qquad e^{10\varphi}\,=\,2^{1/3}X\,, \notag \\
S^5\,=&\,2^{2/3}\left(3K+\alpha\right)\text{vol}_{AdS_3}\,, \notag \\
F^{12}\,=&\,\frac{1}{g}\frac{d}{dy}\left(\frac{y-\alpha}{y+3K_3}\right)dy\wedge\,dz+\frac{1}{g}\text{vol}_{\Sigma_{\mathfrak{g}}}\,, \notag \\
F^{34}\,=&\,\frac{2}{g}\frac{d}{dy}\left(\frac{y-\alpha}{y+3K_1}\right)dy\wedge\,dz\,,
\end{align}
where $\Sigma_{\mathfrak{g}}$ is a Riemann surface and we define
\begin{equation}
H\,=\,\left(y+3K_1\right)^2\left(y+3K_3\right)\,, \qquad P\,=\,H-\left(y-\alpha\right)^2\,, \qquad X=X^{(1)}=X^{(2)}\,=\frac{H^{1/3}}{y+3K_1}\,,
\end{equation}
and $g^2L_{AdS_5}^2=2^{4/3}$. The gauge coupling and the radius of asymptotic $AdS_5$ are fixed to be $g=2^{2/3}$ and $L_{AdS_5}=1$, respectively.

The flux quantization through the Riemann surface is given by
\begin{align}
\mathfrak{s}_1\,=&\,\frac{g}{2\pi}\int_{\Sigma_{\mathfrak{g}}}F^{12}\,=\,2\left(1-\mathfrak{g}\right)\,\in\,\mathbb{Z}\,, \notag \\
\mathfrak{s}_2\,=&\,\frac{g}{2\pi}\int_{\Sigma_{\mathfrak{g}}}F^{34}\,=\,0\,,
\end{align}
where we find $\mathfrak{s}_1+\mathfrak{s}_2=2\left(1-\mathfrak{g}\right)$. Fluxes through the spindle are quantized to be
\begin{align}
\mathfrak{n}_1\,\equiv&\,-\frac{g}{2\pi}\int_{\mathbbl{\Sigma}}F^{12}\,=\,-\frac{\left(y_2-y_1\right)\left(\alpha+3K_3\right)}{\left(y_1+3K_3\right)\left(y_2+3K_3\right)}\frac{\Delta{z}}{2\pi}\,=\,-\frac{p_3}{n_-n_+}\,, \notag \\
2\mathfrak{n}_2\,\equiv&\,-\frac{g}{2\pi}\int_{\mathbbl{\Sigma}}F^{34}\,=\,-2\frac{\left(y_2-y_1\right)\left(\alpha+3K_1\right)}{\left(y_1+3K_1\right)\left(y_2+3K_1\right)}\frac{\Delta{z}}{2\pi}\,=\,-2\frac{p_1}{n_-n_+}\,,
\end{align}
where $p_1$ and $p_3$ are introduced in \eqref{defp5} and $p_i\in\mathbb{Z}$. The minus signs in the definition of $\mathfrak{n}_i$ are introduced for later convenience in the gravitational block calculations. By \eqref{rsymm5} the total flux is obtained to be
\begin{equation}
\mathfrak{n}_1+2\mathfrak{n}_2\,=\,-\frac{2p_1+p_3}{n_-n_+}\,=\,\frac{\eta_1n_+-\eta_2n_-}{n_-n_+}\,,
\end{equation}
where $\eta_1$ and $\eta_2$ are given in \eqref{etaeta} and, thus, both twist and anti-twist solutions are allowed.

\subsection{Uplift to eleven-dimensional supergravity}

We review the uplift formula, \cite{Cvetic:2000ah}, of $U(1)^2$-gauged supergravity in seven dimensions to eleven-dimensional supergravity, \cite{Cremmer:1978km}, as presented in \cite{Cheung:2022ilc}. The metric is given by
\begin{align}
L^{-2}ds_{11}^2\,=\,\Delta^{1/3}ds_7^2+\frac{1}{g^2}\Delta^{-2/3}\left\{e^{4\lambda_1+4\lambda_2}dw_0^2+e^{-2\lambda_1}\left[dw_1^2+w_1^2\left(d\chi_1-gA^{12}\right)^2\right]\right. \notag \\
+\left.e^{-2\lambda_2}\left[dw_2^2+w_2^2\left(d\chi_2-gA^{34}\right)^2\right]\right\}\,,
\end{align}
where
\begin{equation}
\Delta\,=\,e^{-4\lambda_1-4\lambda_2}dw_0^2+e^{2\lambda_1}w_1^2+e^{2\lambda_2}w_2^2\,,
\end{equation}
and $L$ is a length scale. We employ the parametrizations of coordinates of internal four-sphere by
\begin{equation}
\mu^1+i\mu^2\,=\,\cos\xi\cos\theta\,e^{i\chi_1}\,, \qquad \mu^3+i\mu^4\,=\,\cos\xi\sin\theta\,e^{i\chi_2}\,, \qquad \mu^5\,=\,\sin\xi\,,
\end{equation}
with
\begin{equation}
w_0\,=\,\sin\xi\,, \qquad w_1\,=\,\cos\xi\cos\theta\,, \qquad w_2\,=\,\cos\xi\sin\theta\,,
\end{equation}
where $w_0^2+w_1^2+w_2^2=1$ and $\xi\in\left[-\pi/2,\pi/2\right]$, $\theta\in\left[0,\pi/2\right]$, $\chi_1,\,\chi_2\in\left[0,2\pi\right]$.
The four-form flux is given by
\begin{align}
L^{-3}F_{(4)}\,=\,&\frac{w_1w_2}{g^3w_0}U\Delta^{-2}dw_1\wedge\,dw_2\wedge\left(d\chi_1-gA^{12}\right)\wedge\left(d\chi_2-gA^{34}\right) \notag \\
&+\frac{2w_1^2w_2^2}{g^3}\Delta^{-2}e^{2\lambda_1+2\lambda_2}\left(d\lambda_1-d\lambda_2\right)\wedge\left(d\chi_1-gA^{12}\right)\wedge\left(d\chi_2-gA^{34}\right)\wedge\,dw_0 \notag \\
&+\frac{2w_0w_1w_2}{g^3}\Delta^{-2}\left[e^{-2\lambda_1-4\lambda_2}\wedge\left(3d\lambda_1+2d\lambda_2\right)-e^{-4\lambda_1-2\lambda_2}w_2dw_1\wedge\left(2d\lambda_1+3\lambda_2\right)\right] \notag \\
& \qquad \qquad \qquad \qquad \wedge\left(d\chi_1-gA^{12}\right)\wedge\left(d\chi_2-gA^{34}\right) \notag \\
&+\frac{1}{g^2}\Delta^{-1}F^{12}\wedge\left[w_0w_2e^{-4\lambda_1-4\lambda_2}dw_2-w_2^2e^{2\lambda_2}dw_0\right]\wedge\left(d\chi_2-gA^{34}\right) \notag \\
&+\frac{1}{g^2}\Delta^{-1}F^{34}\wedge\left[w_0w_1e^{-4\lambda_1-4\lambda_2}dw_1-w_1^2e^{2\lambda_1}dw_0\right]\wedge\left(d\chi_1-gA^{12}\right) \notag \\
&-w_0e^{-4\lambda_1-4\lambda_2}*_7S^5+\frac{1}{g}S^5\wedge\,dw_0\,,
\end{align}
where
\begin{align}
U\,=\,&\left(e^{-8\lambda_1-8\lambda_2}-2e^{-2\lambda_1-4\lambda_2}-2e^{-4\lambda_1-2\lambda_2}\right)w_0^2 \notag \\
&-\left(e^{-2\lambda_1-4\lambda_2}+2e^{2\lambda_1+2\lambda_2}\right)w_1^2-\left(e^{-4\lambda_1-2\lambda_2}+2e^{2\lambda_1+2\lambda_2}\right)w_2^2\,,
\end{align}
and $*_7$ is a Hodge dual in seven dimensions.

We find a quantization condition of four-form flux through the internal four-sphere,
\begin{align}
\frac{1}{\left(2\pi{l}_p\right)^3}\int_{S^4}F_{(4)}\,=&\,\frac{L^3}{\left(2\pi{l}_p\right)^3}\int_{S^4}\frac{w_1w_2}{g^3w_0}U\Delta^{-2}dw_1\wedge\,dw_2\wedge\,d\chi_1\wedge\,d\chi_2 \notag \\
=&\,\frac{L^3}{\pi{g}^3{l}_p^3}\,\equiv\,N\,\in\,\mathbb{Z}\,,
\end{align}
where $l_p$ is the Planck length and $N$ is the number of M5-branes wrapping $\mathbbl{\Sigma}\times\Sigma_{\mathfrak{g}}$.

For the metric of the form,
\begin{equation}
ds_{11}^2\,=\,e^{2A}\left(ds_{AdS_3}^2+ds_{M_8}^2\right)\,,
\end{equation}
the central charge of dual two-dimensional conformal field theory is given by \cite{Brown:1986nw, Henningson:1998gx}, and we follow \cite{Boido:2021szx},
\begin{equation}
c\,=\,\frac{3}{2G_N^{(3)}}\,=\,\frac{3}{2G_N^{(11)}}\int_{M_8}e^{9A}\text{vol}_{M_8}\,,
\end{equation}
where the eleven-dimensional Newton's gravitational constant is $G_N^{(11)}=\frac{\left(2\pi\right)^8l_p^9}{16\pi}$. For the solutions, with \eqref{ads7y1y2}, we find the holographic central charge to be
\begin{align} \label{holcc}
c\,=&\,\frac{L^9\Delta{z}}{8\pi^5g^6l_p^9}\left(y_2-y_1\right)vol_{\Sigma_{\mathfrak{g}}}\,=\,\frac{\Delta{z}}{2\pi^2}N^3\left(y_1-y_2\right)vol_{\Sigma_{\mathfrak{g}}} \notag \\
=&\,\frac{4q^2\left(n_--n_+-2q\right)}{n_-n_+\left[n_-\left(n_++2q\right)-q\left(2n_++3q\right)\right]}\left(\mathfrak{g}-1\right)N^3\,,
\end{align}
where $vol_{\Sigma_{\mathfrak{g}}}=4\pi\left(\mathfrak{g}-1\right)$. This precisely matches the result obtained from the solutions by uplifting $AdS_3\times\mathbbl{\Sigma}$ to eleven-dimensional supergravity, \cite{Boido:2021szx}.

\section{D4-branes wrapped on $\mathbbl{\Sigma}\times\Sigma_{\mathfrak{g}}$} \label{sec3}

\subsection{Matter coupled $F(4)$ gauged supergravity}

We review $F(4)$ gauged supergravity, \cite{Romans:1985tw}, coupled to a vector multiplet in six dimensions, \cite{Andrianopoli:2001rs, Karndumri:2015eta}, in the conventions of \cite{Faedo:2021nub}. The bosonic field content is consist of the metric, two $U(1)$ gauge fields, $A_i$, a two-form field, $B$, and two scalar fields, $\varphi_i$, where $i\,=\,1,2$. We introduce a parametrization of the scalar fields,
\begin{equation}
X_i\,=\,e^{-\frac{1}{2}\vec{a}_i\cdot\vec{\varphi}}\,, \qquad \vec{a}_1\,=\,\left(2^{1/2},2^{-1/2}\right)\,, \qquad \vec{a}_2\,=\,\left(-2^{1/2},2^{-1/2}\right)\,,
\end{equation}
with
\begin{equation}
X_0\,=\,\left(X_1X_2\right)^{-3/2}\,.
\end{equation}
The field strengths of the gauge fields and two-form field are, respectively,
\begin{equation}
F_i\,=\,dA_i\,, \qquad H=dB\,.
\end{equation}
The action is given by
\begin{align} \label{actionaction}
S\,&=\,\frac{1}{16\pi{G}_N^{(6)}}\int{d}^6x\sqrt{-g}\left[R-V-\frac{1}{2}|d\vec{\varphi}|^2-\frac{1}{2}\sum_{i=1}^2X_i^{-2}|F_i|^2-\frac{1}{8}\left(X_1X_2\right)^2|H|^2\right. \notag \\
&\left.-\frac{m^2}{4}\left(X_1X_2\right)^{-1}|B|^2-\frac{1}{16}\frac{\varepsilon^{\mu\nu\rho\sigma\tau\lambda}}{\sqrt{-g}}B_{\mu\nu}\left(F_{1\rho\sigma}F_{2\tau\lambda}+\frac{m^2}{12}B_{\rho\sigma}B_{\tau\lambda}\right)\right]\,,
\end{align}
where the scalar potential is
\begin{equation}
V\,=\,m^2X_0^2-4g^2X_1X_2-4gmX_0\left(X_1+X_2\right)\,,
\end{equation}
and $\varepsilon_{012345}=+1$. The norm of form fields are defined by
\begin{equation}
|\omega|^2\,=\,\frac{1}{p!}\omega_{\mu_1\ldots\mu_p}\omega^{\mu_1\ldots\mu_p}\,.
\end{equation}
The equations of motion are presented in appendix \ref{appA}.

\subsection{Multi-charged $AdS_2\times\mathbbl{\Sigma}$ solutions} \label{ads2sig}

We review the $AdS_2\times\mathbbl{\Sigma}$ solutions of $U(1)^4$-gauged $\mathcal{N}=2$ supergravity in four dimensions, \cite{Couzens:2021rlk, Ferrero:2021etw}. These solution are obtained from M2-branes wrapped on a spindle, $\mathbbl{\Sigma}$. The metric, gauge fields and scalar fields read
\begin{align} \label{4dbh}
ds_4^2\,=&\,H^{1/2}\left[\frac{1}{4}ds_{AdS_2}^2+\frac{1}{P}dy^2+\frac{P}{4H}dz^2\right]\,, \notag \\
A^{(I)}\,=&\,\frac{y}{y+q_I}dz\,, \qquad X^{(I)}\,=\,\frac{H^{1/4}}{y+q_I}\,,
\end{align}
where $I=1,\ldots,4$ and the functions are defined to be
\begin{equation}
H\,=\,\left(y+q_1\right)\left(y+q_2\right)\left(y+q_3\right)\left(y+q_4\right)\,, \qquad P\,=\,H-4y^2\,.
\end{equation}

In the case of four distinct roots, $y_0<y_1<y_2<y_3$, of quartic polynomial, $P(y)$, the solution is positive and regular in $y\in\left[y_1,y_2\right]$.  The spindle, $\mathbbl{\Sigma}$, is an orbifold, $\mathbbl{WCP}_{[n_1,n_2]}^1$, with conical deficit angles at $y\,=\,y_1,\,y_2$, \cite{Couzens:2021rlk, Ferrero:2021etw}. The spindle numbers, $n_1$, $n_2$, are arbitrary coprime positive integers. The Euler number of the spindle is given by
\begin{equation}
\chi(\mathbbl{\Sigma})\,=\,\frac{1}{4\pi}\int_{\mathbbl{\Sigma}}R_{\mathbbl{\Sigma}}\text{vol}_{\mathbbl{\Sigma}}\,=\,\frac{n_1+n_2}{n_1n_2}\,,
\end{equation}
where $R_{\mathbbl{\Sigma}}$ and $\text{vol}_{\mathbbl{\Sigma}}$ are the Ricci scalar and the volume form on the spindle. The magnetic flux through the spindle is given by
\begin{equation} \label{defp4}
Q_I\,=\,\frac{1}{2\pi}\int_{\mathbbl{\Sigma}}F^{(I)}\,=\,\left(\frac{y_2}{y_2+q_I}-\frac{y_1}{y_1+q_I}\right)\frac{\Delta{z}}{2\pi}\,\equiv\,\frac{2p_I}{n_1n_2}\,,
\end{equation}
and we demand $p_I\in\mathbb{Z}$. One can show that the R-symmetry flux is given by
\begin{equation} \label{rsymm4}
Q^R\,=\,\frac{1}{2}\left(Q_1+Q_2+Q_3+Q_4\right)\,=\,\frac{p_1+p_2+p_2+p_4}{n_1n_2}\,=\,\frac{\eta_1n_2-\eta_2n_1}{n_1n_2}\,,
\end{equation}
where the supersymmetry is realized by, \cite{Ferrero:2021etw, Couzens:2021cpk},
\begin{align}
\text{Anti-twist}: \qquad (\eta_1,\eta_2)\,=\,(+1,+1)\,, \notag \\
\text{Twist}: \qquad \qquad (\eta_1,\eta_2)\,=\,(\pm1,\mp1)\,.
\end{align}
When parameters, $q_I$, $I=1,\ldots,4$, are all identical or identical in pairwise, only the anti-twist solutions are allowed. Otherwise, for all distinct or three identical with one distinct parameters, both the twist and anti-twist solutions are allowed.

Unlike five-dimensional $U(1)^3$-gauged supergravity which has a unique $U(1)^2$ subtruncation, there are two distinct $U(1)^2$ subtruncations from four-dimensional $U(1)^4$-gauged supergravity,
\begin{align}
\text{ST}^2\,\,\text{model}: \qquad A^{(1)}=A^{(2)}\ne{A}^{(3)}=A^{(4)}\,, \qquad X^{(1)}=X^{(2)}\ne{X}^{(3)}=X^{(4)}\,, \notag \\
\text{T}^3\,\,\text{model}: \qquad A^{(1)}=A^{(2)}=A^{(3)}\ne{A}^{(4)}\,, \qquad X^{(1)}=X^{(2)}=X^{(3)}\ne{X}^{(4)}\,,
\end{align}
and their permutations.

\subsection{Multi-charged $AdS_2\times\mathbbl{\Sigma}\times\Sigma_{\mathfrak{g}}$ solutions}

A consistent reduction of matter coupled $F(4)$ gauged supergravity on a Riemann surface was performed in \cite{Hosseini:2020wag}. Inspired by the consistent truncation in \cite{Hosseini:2020wag}, the $AdS_3\times\mathbbl{\Sigma}\times\Sigma_{\mathfrak{g}}$ solutions in \eqref{ads3sigsigsol}, and the minimal $AdS_2\times\mathbbl{\Sigma}\times\Sigma_{\mathfrak{g}}$ solutions in \cite{Faedo:2021nub}, we construct the $AdS_2\times\mathbbl{\Sigma}\times\Sigma_{\mathfrak{g}}$ solutions. However, only the T$^3$ model is obtained from the truncation of $F(4)$ gauged supergravity and not the ST$^2$ model. Thus, we only find solutions by uplifting multi-charged $AdS_2\times\mathbbl{\Sigma}$ solutions in section \ref{ads2sig} with
\begin{equation}
q_1=q_2=q_3\ne{q}_4\,,
\end{equation}
to six dimensions. After some trial and error we find the solutions to be
\begin{align}
ds_6^2\,=&\,e^{-2C}L_{AdS_4}^2H^{1/2}\left[\frac{1}{4}ds_{AdS_2}^2+\frac{1}{P}dy^2+\frac{P}{4H}dz^2\right]+e^{2C}ds_{\Sigma_{\mathfrak{g}}}^2\,, \notag \\
X_1\,=&\,k_8^{1/8}k_2^{1/2}\frac{H^{1/4}}{y+q_1}\,, \qquad X_2\,=\,k_8^{1/8}k_2^{-1/2}\frac{H^{1/4}}{y+q_1}\,, \qquad e^{-2C}\,=\,m^2k_8^{1/4}k_4\frac{H^{1/4}}{y+q_1} \notag \\
B\,=&\,q_1\frac{9k_8^{1/2}}{4g^2}\text{vol}_{AdS_2}\,,\notag \\
F_1\,=&\,\frac{3k_8^{1/2}k_2^{1/2}}{2g}\frac{q_1}{\left(y+q_1\right)^2}dy\wedge\,dz+\frac{\kappa+\mathtt{z}}{2g}\text{vol}_{\Sigma_{\mathfrak{g}}}\,, \notag \\
F_2\,=&\,\frac{3k_8^{1/2}k_2^{-1/2}}{2g}\frac{q_4}{\left(y+q_4\right)^2}dy\wedge\,dz+\frac{\kappa-\mathtt{z}}{2g}\text{vol}_{\Sigma_{\mathfrak{g}}}\,,
\end{align}
where we define
\begin{equation}
H\,=\,\left(y+q_1\right)^3\left(y+q_4\right)\,, \qquad P\,=\,H-4y^2\,,
\end{equation}
and
\begin{equation}
g\,=\,\frac{3m}{2}\,, \qquad L_{AdS_4}\,=\,\frac{k_8^{1/4}k_4^{-1/2}}{m^2}\,.
\end{equation}
There are parameters, $\kappa=0,\,\pm1$, for the curvature of Riemann surface, and, $\mathtt{z}$, which define
\begin{equation}
k_2\,=\,\frac{3\mathtt{z}+\sqrt{\kappa^2+8\mathtt{z}^2}}{\mathtt{z}-\kappa}\,, \qquad k_8\,=\,\frac{16k_2}{9\left(1+k_2\right)^2}\,, \qquad k_4\,=\,\frac{18}{-3\kappa+\sqrt{\kappa^2+8\mathtt{z}^2}}\,.
\end{equation}
If we set $q_1=q_2=q_3=q_4$, it reduces to the minimal $AdS_2\times\mathbbl{\Sigma}\times\Sigma_{\mathfrak{g}}$ solutions in \cite{Faedo:2021nub}. For our solutions, in order to satisfy the equations of motion, we find that we should choose
\begin{equation} \label{fixzz}
\kappa\,=\,-1\,, \qquad \mathtt{z}\,=\,1\,,
\end{equation}
and we find $k_2=k_4=k_8^{-1}=3$. Then the solutions are given by
\begin{align}
ds_6^2\,=&\,e^{-2C}L_{AdS_4}^2H^{1/2}\left[\frac{1}{4}ds_{AdS_2}^2+\frac{1}{P}dy^2+\frac{P}{4H}dz^2\right]+e^{2C}ds_{\Sigma_{\mathfrak{g}}}^2\,, \notag \\
X_1\,=&\,3^{3/8}\frac{H^{1/4}}{y+q_1}\,, \qquad X_2\,=\,3^{-5/8}\frac{H^{1/4}}{y+q_1}\,, \qquad e^{-2C}\,=\,\frac{4g^2}{3^{5/4}}\frac{H^{1/4}}{y+q_1} \notag \\
B\,=&\,q_1\frac{3\sqrt{3}}{4g^2}\text{vol}_{AdS_2}\,,\notag \\
F_1\,=&\,\frac{3}{2g}\frac{q_1}{\left(y+q_1\right)^2}dy\wedge\,dz\,, \notag \\
F_2\,=&\,\frac{1}{2g}\frac{q_4}{\left(y+q_4\right)^2}dy\wedge\,dz-\frac{1}{g}\text{vol}_{\Sigma_{\mathfrak{g}}}\,,
\end{align}
where we have
\begin{equation}
g\,=\,\frac{3m}{2}\,, \qquad L_{AdS_4}\,=\,\frac{3^{5/4}}{4g^2}\,.
\end{equation}
Notice that the components of $F_1$ on the Riemann surface is turned off by the choice of \eqref{fixzz}.

The flux quantization through the Riemann surface is given by
\begin{align}
\mathfrak{s}_1\,=&\,\frac{g}{2\pi}\int_{\Sigma_{\mathfrak{g}}}F_1\,=\,0\,, \notag \\
\mathfrak{s}_2\,=&\,\frac{g}{2\pi}\int_{\Sigma_{\mathfrak{g}}}F_2\,=\,2\left(1-\mathfrak{g}\right)\,\in\,\mathbb{Z}\,,
\end{align}
where we find $\mathfrak{s}_1+\mathfrak{s}_2=2\left(1-\mathfrak{g}\right)$. Fluxes through the spindle are quantized to be
\begin{align}
3\mathfrak{n}_1\,\equiv\,\frac{g}{2\pi}\int_{\mathbbl{\Sigma}}F_1\,=\,\frac{3}{2}\left(\frac{y_2}{y_2+q_1}-\frac{y_1}{y_1+q_1}\right)\frac{\Delta{z}}{2\pi}\,=\,\frac{3p_1}{n_1n_2}\,, \notag \\
\mathfrak{n}_2\,\equiv\,\frac{g}{2\pi}\int_{\mathbbl{\Sigma}}F_2\,=\,\frac{1}{2}\left(\frac{y_2}{y_2+q_4}-\frac{y_1}{y_1+q_4}\right)\frac{\Delta{z}}{2\pi}\,=\,\frac{p_4}{n_1n_2}\,,
\end{align}
where $p_1$ and $p_4$ are introduced in \eqref{defp4} and $p_i\in\mathbb{Z}$. By \eqref{rsymm4} the total flux is obtained to be
\begin{equation}
3\mathfrak{n}_1+\mathfrak{n}_2\,=\,\frac{3p_1+p_4}{n_1n_2}\,=\,\frac{\eta_1n_2-\eta_2n_1}{n_1n_2}\,,
\end{equation}
and both the twist and anti-twist solutions are allowed.{\footnote{We would like to thank Chris Couzens for discussion on this.}}

\subsection{Uplift to massive type IIA supergravity}

We review the uplift formula, \cite{Cvetic:1999xx}, of matter coupled $F(4)$ gauged supergravity to massive type IIA supergravity, \cite{Romans:1985tz}, presented in \cite{Faedo:2021nub}. Although the uplift formula is only given for vanishing of two-form field, $B$, in $F(4)$ gauged supergravity, it correctly reproduces the metric, the dilaton and the internal four-sphere part of four-form flux. The metric in the string frame and the dilaton field are
\begin{align}
ds^2_{\text{s.f.}}\,&=\,\lambda^2\mu_0^{-1/3}\left(X_1X_2\right)^{-1/4}\left\{\Delta^{1/2}ds_6^2\right. \notag \\
&\left.+g^{-2}\Delta^{-1/2}\left[X_0^{-1}d\mu_0^2+X_1^{-1}\left(d\mu_1^2+\mu_1^2\sigma_1^2\right)+X_2^{-1}\left(d\mu_2^2+\mu_2^2\sigma_2^2\right)\right]\right\}\,, \\
e^\Phi\,&=\,\lambda^2\mu_0^{-5/6}\Delta^{1/4}\left(X_1X_2\right)^{-5/8}\,,
\end{align}
where the function, $\Delta$, is defined by
\begin{equation}
\Delta\,=\,\sum_{a=0}^2X_a\mu_a^2\,,
\end{equation}
and the one-forms are $\sigma_i=d\phi_i-gA_i$. The angular coordinates, $\phi_1$, $\phi_2$, have canonical periodicities of $2\pi$. We employ the parametrization of coordinates,
\begin{equation}
\mu_0\,=\,\sin\xi\,, \qquad \mu_1\,=\,\cos\xi\sin\eta\,, \qquad \mu_2\,=\,\cos\xi\cos\eta\,,
\end{equation}
where $\sum_{a=0}^2\mu_a^2=1$ and $\eta\in[0,\pi/2]$, $\xi\in(0,\pi/2]$. The internal space is a squashed four-hemisphere which has a singularity on the boundary, $\xi\rightarrow0$. The four-form flux is given by
\begin{equation}
\lambda^{-1}*F_{(4)}\,=\,gU\text{vol}_6-\frac{1}{g^2}\sum_{i=1}^2X_i^{-2}\mu_i\left(*_6F_i\right)\wedge{d}\mu_i\wedge\sigma_i+\frac{1}{g}\sum_{a=0}^2X_a^{-1}\mu_a\left(*_6dX_a\right)\wedge{d}\mu_a\,,
\end{equation}
where the function, $U$, is defined by
\begin{equation}
U\,=\,2\sum_{a=0}^2X_a^2\mu_a^2-\left[\frac{4}{3}X_0+2\left(X_1+X_2\right)\right]\Delta\,,
\end{equation}
and $*_6$ is a Hodge dual in six dimensions. The Romans mass is given by
\begin{equation}
F_{(0)}\,=\,\frac{2g}{3\lambda^3}\,.
\end{equation}
The positive constant, $\lambda$, is introduced from the scaling symmetry of the theory. It plays an important role to have regular solutions with proper flux quantizations, \cite{Faedo:2021nub}. The uplift formula implies $m=2g/3$. 

The relevant part of the four-form flux for flux quantization is the component on the internal four-sphere,
\begin{align}
F_{(4)}\,=&\,\frac{\lambda\mu_0^{1/3}}{g^3\Delta}\frac{U}{\Delta}\frac{\mu_1\mu_2}{\mu_0}d\mu_1\wedge{d}\mu_2\wedge\sigma_1\wedge\sigma_2+\ldots\,.
\end{align}
We impose quantization conditions on the fluxes,
\begin{equation}
\left(2\pi{l}_s\right)F_{(0)}\,=\,n_0\,\in\,\mathbb{Z}\,, \qquad \frac{1}{\left(2\pi{l}_s\right)^3}\int_{\tilde{S}^4}F_{(4)}\,=\,N\,\in\,\mathbb{Z}\,,
\end{equation}
where $l_s$ is the string length. For the solutions, these imply that
\begin{equation}
g^8\,=\,\frac{1}{\left(2\pi{l}_s\right)^8}\frac{18\pi^6}{N^3n_0}\,, \qquad \lambda^8\,=\,\frac{8\pi^2}{9Nn_0^3}\,,
\end{equation}
where we have $n_0=8-N_f$ and $N_f$ is the number of D8-branes. These results are identical to the case of minimal $AdS_2\times\mathbbl{\Sigma}\times\Sigma_{\mathfrak{g}}$ solutions in \cite{Faedo:2021nub}.

For the metric of the form in the string frame,
\begin{equation}
ds_{\text{s.f.}}^2\,=\,e^{2A}\left(ds_{AdS_2}^2+ds_{M_8}^2\right)\,,
\end{equation}
the Bekenstein-Hawking entropy is by, \cite{Brown:1986nw, Henningson:1998gx}, and in \cite{Faedo:2021nub},
\begin{equation} \label{bhenm8}
S_{\text{BH}}\,=\,\frac{1}{4G_N^{(2)}}\,=\,\frac{8\pi^2}{\left(2\pi{l}_s\right)^8}\int{e}^{8A-2\Phi}\text{vol}_{M_8}\,.
\end{equation}
For the solutions, we obtain the Bekenstein-Hawking entropy to be
\begin{equation}
S_{\text{BH}}\,=\,\frac{1}{\left(2\pi{l}_s\right)^8}\frac{9\left(3\pi\lambda\right)^4k_8^{1/2}}{20g^8k_4}4\pi\kappa\left(1-\mathfrak{g}\right)A_h\,=\,\frac{1}{\left(2\pi{l}_s\right)^8}\frac{\sqrt{3}\left(3\pi\lambda\right)^4}{20g^8}4\pi\kappa\left(1-\mathfrak{g}\right)A_h\,,
\end{equation}
where the area of the horizon of black hole, multi-charged $AdS_2\times\mathbbl{\Sigma}$, in \eqref{4dbh} is
\begin{equation}
A_h\,=\,\frac{1}{2}\left(y_2-y_1\right)\Delta{z}\,,
\end{equation}
and $y_1$ and $y_2$ are two relevant roots of $P(y)$. The free energy of 5d $USp(2N)$ gauge theory on $S^3\times\,\Sigma_{\mathfrak{g}}$ is given by, \cite{Bah:2018lyv, Crichigno:2018adf, Faedo:2021nub},
\begin{align} \label{free3s}
\mathcal{F}_{S^3\times\,\Sigma_{\mathfrak{g}}}\,=&\,\frac{16\pi^3}{\left(2\pi\,l_s\right)^8}\int\,e^{8A-2\Phi}\text{vol}_{M_6} \notag \\
=&\,\frac{16\pi\kappa\left(1-\mathfrak{g}\right)N^{5/2}\left(\mathtt{z}^2-\kappa^2\right)^{3/2}\left(\sqrt{\kappa^2+8\mathtt{z}^2}-\kappa\right)}{5\sqrt{8-N_f}\left(\kappa\sqrt{\kappa^2+8\mathtt{z}^2}-\kappa^2+4\mathtt{z}^2\right)^{3/2}}\,.
\end{align}
By comparing \eqref{free3s} with \eqref{bhenm8}, we find the Bekenstein-Hawking entropy to be
\begin{align}
S_{\text{BH}}\,=&\,\frac{1}{2\pi}\mathcal{F}_{S^3\times\,\Sigma_{\mathfrak{g}}}A_h \notag \\
=&\,\frac{8\kappa\left(1-\mathfrak{g}\right)N^{5/2}\left(\mathtt{z}^2-\kappa^2\right)^{3/2}\left(\sqrt{\kappa^2+8\mathtt{z}^2}-\kappa\right)}{5\sqrt{8-N_f}\left(\kappa\sqrt{\kappa^2+8\mathtt{z}^2}-\kappa^2+4\mathtt{z}^2\right)^{3/2}}A_h\,,
\end{align}
and, for $\kappa=-1$ and $\mathtt{z}=1$, \eqref{fixzz}, we obtain{\footnote{We would like to thank Hyojoong Kim for comments on this limit.}}
\begin{equation}
S_{\text{BH}}\,=\,\left(\frac{3}{8}\right)^{3/2}\frac{32\left(\mathfrak{g}-1\right)N^{5/2}}{5\sqrt{8-N_f}}A_h\,.
\end{equation}
 Although formally the Bekenstein-Hawking entropy is in the identical expression of the one for minimal $AdS_2\times\mathbbl{\Sigma}\times\Sigma_{\mathfrak{g}}$ solutions in \cite{Faedo:2021nub}, note that the black holes that give the area, $A_h$, are different: it was minimal $AdS_2\times\mathbbl{\Sigma}$ in \cite{Faedo:2021nub}, but now it is multi-charged $AdS_2\times\mathbbl{\Sigma}$, \cite{Ferrero:2021etw}. We refer \cite{Couzens:2021cpk} for the explicit expression of $A_h$ for the multi-charged solutions.

\section{Gravitational blocks} \label{sec4}

In this section, we briefly review the off-shell quantities from gluing gravitational blocks, \cite{Hosseini:2019iad}, and show that extremization of off-shell quantity correctly reproduces the Bekenstein-Hawking entropy, central charge, and free energy, depending on the dimensionality, \cite{Faedo:2021nub}. Then apply the gravitational block calculations to the solutions we constructed in the previous sections.

Depending on the dimensionality, the Bekenstein-Hawking entropy, central charge, and free energy are obtained by extremizing the off-shell quantity, \cite{Faedo:2021nub},
\begin{equation} \label{fpmd}
F_d^\pm\left(\Delta_i,\epsilon;\mathfrak{n}_i,n_+,n_-,\sigma\right)\,=\,\frac{1}{\epsilon}\Big(\mathcal{F}_d\left(\Delta_i^+\right)\pm\mathcal{F}_d\left(\Delta_i^-\right)\Big)\,,
\end{equation}
where $\mathcal{F}_d$ are the gravitational blocks, \cite{Hosseini:2019iad}. We also define quantities,
\begin{equation}
\Delta^\pm_i\,\equiv\,\varphi_i\pm\mathfrak{n}_i\epsilon\,,
\end{equation}
and
\begin{equation}
\varphi_i\,\equiv\,\Delta_i+\frac{r_i}{2}\frac{n_+-\sigma{n}_-}{n_+n_-}\epsilon\,,
\end{equation}
where $\sigma=+1$ and $\sigma=-1$ for twist and anti-twist solutions, respectively. The expressions of gravitational blocks are 
\begin{equation}
\mathcal{F}_3\,=\,b_3\left(\Delta_1\Delta_2\Delta_3\Delta_4\right)^{1/2}\,, \quad \mathcal{F}_4\,=\,b_4\left(\Delta_1\Delta_2\Delta_3\right)\,, \quad \mathcal{F}_5\,=\,b_5\left(\Delta_1\Delta_2\right)^{3/2}\,, \quad \mathcal{F}_6\,=\,b_6\left(\Delta_1\Delta_2\right)^2\,,
\end{equation}
and the constants, $b_d$, will be given later. The relative sign for gluing gravitational blocks in \eqref{fpmd} is $-\sigma$ for $d=3,5$ and $-$ for $d=4,6$. The twist conditions on the magnetic flux through the spindle, $\mathfrak{n}_i$, is given by
\begin{equation} \label{blocktwist}
\sum_{i=1}^\mathfrak{d}\mathfrak{n}_i\,=\,\frac{n_++\sigma{n}_-}{n_+n_-}\,,
\end{equation}
where $n_+$ and $n_-$ are the orbifold numbers of spindle and $\mathfrak{d}$ is the rank of global symmetry group of dual field theory, $i.e.$, $\mathfrak{d}=4$ for $d=3$, $\mathfrak{d}=3$ for $d=4$, and $\mathfrak{d}=2$ for $d=5,6$. The constants are constrained by
\begin{equation} \label{flanorm}
\sum_{i=1}^\mathfrak{d}r_i\,=\,2\,,
\end{equation}
and they parametrize the ambiguities of defining the flavor symmetries. The $U(1)$ R-symmetry flux gives
\begin{equation}
\frac{1}{2\pi}\int_\mathbbl{\Sigma}dA_R\,=\,\frac{n_++\sigma{n}_-}{n_+n_-}\,,
\end{equation}
and the fugacities of dual field theories are normalized by
\begin{equation} \label{fugnorm}
\sum_{i=1}^\mathfrak{d}\Delta_i\,=\,2\,.
\end{equation}
The off-shell quantity can be written by
\begin{equation} \label{offef}
F_d^\pm\left(\varphi_i,\epsilon;\mathfrak{n}\right)\,=\,\frac{1}{\epsilon}\Big(\mathcal{F}_d\left(\varphi_i+\mathfrak{n}_i\epsilon\right)\pm\mathcal{F}_d\left(\varphi_i-\mathfrak{n}_i\epsilon\right)\Big)\,,
\end{equation}
where the variables satisfy the constraint,
\begin{equation} \label{varphicst}
\sum_{i=1}^\mathfrak{d}\varphi_i-\frac{n_+-\sigma{n}_-}{n_+n_-}\epsilon\,=\,2\,,
\end{equation}
which originates from \eqref{flanorm} and \eqref{fugnorm}.

\subsection{M5-branes wrapped on $\mathbbl{\Sigma}\times\Sigma_{\mathfrak{g}}$}

For the $AdS_3\times\mathbbl{\Sigma}\times\Sigma_{\mathfrak{g}}$ solutions, there is standard topological twist on $\Sigma_\mathfrak{g}$ for the magnetic charges, $\mathfrak{s}_i$, and anti-twist on $\mathbbl{\Sigma}$ for $\mathfrak{n}_i$. Then the off-shell central charge is given by
\begin{align}
S(\varphi_i,\epsilon_1,\epsilon_2;\mathfrak{n}_i,\mathfrak{s}_i)\,=\,-\frac{1}{4\epsilon_1\epsilon_2}\Big[&\mathcal{F}_6(\varphi_i+\mathfrak{n}_i\epsilon_1+\mathfrak{s}_i\epsilon_2)-\mathcal{F}_6(\varphi_i-\mathfrak{n}_i\epsilon_1+\mathfrak{s}_i\epsilon_2) \notag \\
-&\mathcal{F}_6(\varphi_i+\mathfrak{n}_i\epsilon_1-\mathfrak{s}_i\epsilon_2)+\mathcal{F}_6(\varphi_i-\mathfrak{n}_i\epsilon_1-\mathfrak{s}_i\epsilon_2)\Big]\,,
\end{align}
with the constraints,
\begin{align}
\mathfrak{n}_1+2\mathfrak{n}_2\,=\,\frac{n_+-n_-}{n_+n_-}\,,\qquad \mathfrak{s}_1+\mathfrak{s}_2\,=\,2(1-\mathfrak{g})\,, \qquad \varphi_1+2\varphi_2-\frac{n_++n_-}{n_+n_-}\epsilon_1\,=\,2\,.
\end{align}
For the calculations, we employ
\begin{equation}
b_4\,=\,-\frac{3}{2}N^2\,, \qquad b_6\,=\,-N^3\,.
\end{equation}
Extremizing it with respect to $\epsilon_2$ gives $\epsilon_2=0$ and renaming $\epsilon_1\mapsto\epsilon$, we find the off-shell central charge expressed by
\begin{align} \label{oscc1}
S(\varphi_i,\epsilon;\mathfrak{n}_i,\mathfrak{s}_i)\,=&\,2N^3\mathfrak{s}_1\left(\mathfrak{n}_1\varphi_2\varphi_3+\varphi_1\mathfrak{n}_2\varphi_3+\varphi_1\varphi_2\mathfrak{n}_3+\mathfrak{n}_1\mathfrak{n}_2\mathfrak{n}_3\epsilon^2\right)|_{3\mapsto2} \notag \\
& \,+2N^3\mathfrak{s}_2\left(\mathfrak{n}_1\varphi_2\varphi_3+\varphi_1\mathfrak{n}_2\varphi_3+\varphi_1\varphi_2\mathfrak{n}_3+\mathfrak{n}_1\mathfrak{n}_2\mathfrak{n}_3\epsilon^2\right)|_{3\mapsto1} \notag \\
=&\,2N^3\mathfrak{s}_1\left.\left(-\frac{1}{3N^2}F_4^-\right)\right|_{3\mapsto2}+2N^3\mathfrak{s}_2\left.\left(-\frac{1}{3N^2}F_4^-\right)\right|_{3\mapsto1}\,.
\end{align}
We have started with the $d=6$ gravitational blocks, $\mathcal{F}_6$, and we observe the $d=4$ structure, $F_4^-$, naturally emerges. See section 5.2 of \cite{Faedo:2021nub} for the calculations of $d=4$ gravitational blocks. From the $d=4$ point of view, the $\mathfrak{s}_1$ term of $S(\varphi_i,\epsilon;\mathfrak{n}_i,\mathfrak{s}_i)$ in \eqref{oscc1} is the off-shell central charge for $\mathfrak{n}_1\ne\mathfrak{n}_2=\mathfrak{n}_3$ and the $\mathfrak{s}_2$ terms is for $\mathfrak{n}_1=\mathfrak{n}_3\ne\mathfrak{n}_2$. Thus, extremization gives disparate results for each term. However, for the solution, as we have 
\begin{equation}
\mathfrak{s}_1\,=\,2\left(1-\mathfrak{g}\right)\,, \qquad \mathfrak{s}_2\,=\,0\,,
\end{equation}
the solution chooses the $\mathfrak{s}_1$ term in the off-shell central charge. Extremizing this we find the values,
\begin{equation}
\epsilon^*\,=\,\left.\frac{\frac{n_+-\sigma{n}_-}{n_+n_-}}{2\left(\frac{\sigma}{n_+n_-}-\left(\mathfrak{n}_1\mathfrak{n}_2+\mathfrak{n}_2\mathfrak{n}_3+\mathfrak{n}_3\mathfrak{n}_1\right)\right)}\right|_{3\mapsto2}, \, \varphi_2^*\,=\,\left.\frac{\mathfrak{n}_2\left(\mathfrak{n}_2-\mathfrak{n}_3-\mathfrak{n}_1\right)}{2\left(\frac{\sigma}{n_+n_-}-\left(\mathfrak{n}_1\mathfrak{n}_2+\mathfrak{n}_2\mathfrak{n}_3+\mathfrak{n}_3\mathfrak{n}_1\right)\right)}\right|_{3\mapsto2}.
\end{equation}
Then the off-shell central charge gives
\begin{align}
S(\varphi_i^*,\epsilon^*;\mathfrak{n}_i)\,=\,4N^3\left(\mathfrak{g}-1\right)\left.\frac{\mathfrak{n}_1\mathfrak{n}_2\mathfrak{n}_3}{\frac{\sigma}{n_+n_-}-\left(\mathfrak{n}_1\mathfrak{n}_2+\mathfrak{n}_2\mathfrak{n}_3+\mathfrak{n}_3\mathfrak{n}_1\right)}\right|_{3\mapsto2}\,,
\end{align}
which precisely matches the holographic central charge from the supergravity solutions, \eqref{holcc}, with $\sigma=-1$.

\subsection{D4-branes wrapped on $\mathbbl{\Sigma}\times\Sigma_{\mathfrak{g}}$}

For the $AdS_2\times\mathbbl{\Sigma}\times\Sigma_{\mathfrak{g}}$ solutions, there is standard topological twist on $\Sigma_\mathfrak{g}$ for the magnetic charges, $\mathfrak{s}_i$, and anti-twist on $\mathbbl{\Sigma}$ for $\mathfrak{n}_i$. Then the entropy function is given by
\begin{align}
S(\varphi_i,\epsilon_1,\epsilon_2;\mathfrak{n}_i,\mathfrak{s}_i)\,=\,-\frac{1}{4\epsilon_1\epsilon_2}\Big[&\mathcal{F}_5(\varphi_i+\mathfrak{n}_i\epsilon_1+\mathfrak{s}_i\epsilon_2)+\mathcal{F}_5(\varphi_i-\mathfrak{n}_i\epsilon_1+\mathfrak{s}_i\epsilon_2) \notag \\
-&\mathcal{F}_5(\varphi_i+\mathfrak{n}_i\epsilon_1-\mathfrak{s}_i\epsilon_2)-\mathcal{F}_5(\varphi_i-\mathfrak{n}_i\epsilon_1-\mathfrak{s}_i\epsilon_2)\Big]\,,
\end{align}
with the constraints,
\begin{align}
\mathfrak{n}_1+3\mathfrak{n}_2\,=\,\frac{n_+-n_-}{n_+n_-}\,,\qquad \mathfrak{s}_1+\mathfrak{s}_2\,=\,2(1-\mathfrak{g})\,, \qquad \varphi_1+3\varphi_2-\frac{n_++n_-}{n_+n_-}\epsilon_1\,=\,2\,.
\end{align}
For the calculations, we employ
\begin{equation}
b_3\,=\,-\frac{\sqrt{2}\pi}{3}N^{3/2}\,, \qquad b_5\,=\,-\frac{2^{5/2}\pi}{15}\frac{N^{5/2}}{\sqrt{8-N_f}}\,.
\end{equation}
Extremizing it with respect to $\epsilon_2$ gives $\epsilon_2=0$ and renaming $\epsilon_1\mapsto\epsilon$, we find the entropy function expressed by
\begin{align} \label{oscc2}
S(\varphi_i,\epsilon;\mathfrak{n}_i,\mathfrak{s}_i)\,=\frac{c}{\epsilon}&\left[\mathfrak{s}_1\left(\sqrt{\left(\varphi_1+\mathfrak{n}_1\epsilon\right)\left(\varphi_2+\mathfrak{n}_2\epsilon\right)^3}+\sqrt{\left(\varphi_1-\mathfrak{n}_1\epsilon\right)\left(\varphi_2-\mathfrak{n}_2\epsilon\right)^3}\right)\right. \notag \\
&\left.+\mathfrak{s}_2\left(\sqrt{\left(\varphi_1+\mathfrak{n}_1\epsilon\right)^3\left(\varphi_2+\mathfrak{n}_2\epsilon\right)}+\sqrt{\left(\varphi_1-\mathfrak{n}_1\epsilon\right)^3\left(\varphi_2-\mathfrak{n}_2\epsilon\right)}\right)\right]\,,
\end{align}
where we have
\begin{equation}
c\,\equiv\,\frac{\sqrt{2}\pi}{5}\frac{N^{5/2}}{\sqrt{8-N_f}}\,.
\end{equation}
We have started with the $d=5$ gravitational blocks, $\mathcal{F}_5$, and we observe the $d=3$ structure naturally emerges. See section 5.1 of \cite{Faedo:2021nub} for the calculations of $d=3$ gravitational blocks. From the $d=3$ point of view, the $\mathfrak{s}_1$ term of $S(\varphi_i,\epsilon;\mathfrak{n}_i,\mathfrak{s}_i)$ in \eqref{oscc2} is the entropy function for $\mathfrak{n}_1\ne\mathfrak{n}_2=\mathfrak{n}_3=\mathfrak{n}_4$ and the $\mathfrak{s}_2$ terms is for $\mathfrak{n}_1=\mathfrak{n}_2=\mathfrak{n}_3\ne\mathfrak{n}_4$. Thus, extremization gives disparate results for each term. However, for the solution, as we have 
\begin{equation}
\mathfrak{s}_1\,=\,2\left(1-\mathfrak{g}\right)\,, \qquad \mathfrak{s}_2\,=\,0\,,
\end{equation}
the solution chooses the $\mathfrak{s}_1$ term in the entropy function. However, in this case, the algebraic equations appearing in the extremization procedure are quite complicated and we do not pursue it further here.

\section{Conclusions} \label{sec5}

In this work, we have constructed multi-charged $AdS_3\times\mathbbl{\Sigma}\times\Sigma_{\mathfrak{g}}$ and $AdS_2\times\mathbbl{\Sigma}\times\Sigma_{\mathfrak{g}}$ solutions from M5-branes and D4-branes. We have uplifted the solutions to eleven-dimensional and massive type IIA supergravity, respectively. We have also studied their spindle properties and calculated the holographic central charge and the Bekenstein-Hawking entropy, respectively.

Although we have only considered the $AdS_{2,3}\times\mathbbl{\Sigma}\times\Sigma_{\mathfrak{g}}$ solutions for spindle, $\mathbbl{\Sigma}$, the local form of our solutions naturally allows solutions for disk, $\mathbbl{\Sigma}$, by different global completion. However, the $AdS_3\times\mathbbl{\Sigma}\times\Sigma_{\mathfrak{g}}$ solution for disk, $\mathbbl{\Sigma}$, was already constructed and studied in \cite{Suh:2021ifj}. Thus, it would be interesting to analyze the $AdS_2\times\mathbbl{\Sigma}\times\Sigma_{\mathfrak{g}}$ solutions for disk, $\mathbbl{\Sigma}$, from the solutions we have constructed.

Unlike the minimal $AdS_2\times\mathbbl{\Sigma}\times\Sigma_{\mathfrak{g}}$ solutions in \cite{Faedo:2021nub} where $\mathtt{z}$ is a free parameter, only $\mathtt{z}=1$ is allowed for our multi-charged $AdS_2\times\mathbbl{\Sigma}\times\Sigma_{\mathfrak{g}}$ solutions, \eqref{fixzz}. We would like to understand why it is required to fix the parameter for the solutions and if there are more general multi-charged solutions with additional parameters.

The solutions we have obtained could be seen as generalizations of $AdS_3\times\Sigma_{\mathfrak{g}_1}\times\Sigma_{\mathfrak{g}_2}$ solutions in \cite{Gauntlett:2001jj} and $AdS_2\times\Sigma_{\mathfrak{g}_1}\times\Sigma_{\mathfrak{g}_2}$ solutions in \cite{Suh:2018tul, Hosseini:2018usu, Suh:2018szn}. In particular, via the AdS/CFT correspondence, \cite{Maldacena:1997re}, the Bekenstein-Hawking entropy of $AdS_2\times\Sigma_{\mathfrak{g}_1}\times\Sigma_{\mathfrak{g}_2}$ solutions was microscopically counted by the topologically twisted index of 5d $USp(2N)$ gauge theories, \cite{Hosseini:2018uzp, Crichigno:2018adf}. It would be most interesting to derive the Bekenstein-Hawking entropy of the $AdS_2\times\mathbbl{\Sigma}\times\Sigma_{\mathfrak{g}}$ solutions from the field theory calculations.

\bigskip
\bigskip
\leftline{\bf Acknowledgements}
\noindent We would like to thank Chris Couzens, Hyojoong Kim, Nakwoo Kim, and Yein Lee for interesting discussions and collaborations in a related project. This research was supported by the National Research Foundation of Korea under the grant NRF-2019R1I1A1A01060811.

\appendix
\section{The equations of motion} \label{appA}
\renewcommand{\theequation}{A.\arabic{equation}}
\setcounter{equation}{0} 

\subsection{$U(1)^2$-gauged supergravity in seven dimensions}

We present the equations of motion  derived from the Lagrangian in \eqref{sevenlag},
\begin{align}
R_{\mu\nu}\,=\,&6\partial_\mu\lambda_1\partial_\nu\lambda_1+6\partial_\mu\lambda_2\partial_\nu\lambda_2+8\partial_{(\mu}\lambda_1\partial_{\nu)}\lambda_2+\frac{1}{5}g_{\mu\nu}V \notag \\
&+\frac{1}{2}e^{-4\lambda_1}\left(F_{\mu\rho}^{12}F_\nu^{12\rho}-\frac{1}{10}g_{\mu\nu}F_{\rho\sigma}^{12}F^{12\rho\sigma}\right)+\frac{1}{2}e^{-4\lambda_2}\left(F_{\mu\rho}^{34}F_\nu^{34\rho}-\frac{1}{10}g_{\mu\nu}F_{\rho\sigma}^{34}F^{34\rho\sigma}\right) \notag \\
&+\frac{1}{4}e^{-4\lambda_1-4\lambda_2}\left(S_{\mu\rho\sigma}^5S_\nu^{5\rho\sigma}-\frac{2}{15}g_{\mu\nu}S_{\rho\sigma\delta}^5S^{5\rho\sigma\delta}\right)\,,
\end{align}
\begin{align}
\frac{1}{\sqrt{-g}}\partial_\mu\left(\sqrt{-g}g^{\mu\nu}\partial_\nu\left(3\lambda_1+2\lambda_2\right)\right)+\frac{1}{4}e^{-4\lambda_1}F_{\mu\nu}^{12}F^{12\mu\nu}+\frac{1}{12}e^{-4\lambda_1-4\lambda_2}S_{\mu\nu\rho}^5S^{5\mu\nu\rho}-\frac{g^2}{4}\frac{\partial{V}}{\partial\lambda_1}\,=\,0\,, \notag \\
\frac{1}{\sqrt{-g}}\partial_\mu\left(\sqrt{-g}g^{\mu\nu}\partial_\nu\left(2\lambda_1+3\lambda_2\right)\right)+\frac{1}{4}e^{-4\lambda_2}F_{\mu\nu}^{34}F^{34\mu\nu}+\frac{1}{12}e^{-4\lambda_1-4\lambda_2}S_{\mu\nu\rho}^5S^{5\mu\nu\rho}-\frac{g^2}{4}\frac{\partial{V}}{\partial\lambda_2}\,=\,0\,, 
\end{align}
\begin{align}
d\left(e^{-4\lambda_1}*F^{12}\right)+e^{-4\lambda_1-4\lambda_2}*S^5\wedge\,F^{34}\,=\,0\,, \notag \\
d\left(e^{-4\lambda_2}*F^{34}\right)+e^{-4\lambda_1-4\lambda_2}*S^5\wedge\,F^{12}\,=\,0\,, \notag \\
dS^5-ge^{-4\lambda_1-4\lambda_2}*S^5-F^{12}\wedge\,F^{34}\,=\,0\,.
\end{align}

\subsection{Matter coupled $F(4)$ gauged supergravity}

We present the equations of motion  derived from the action in \eqref{actionaction},
\begin{align}
&R_{\mu\nu}-\frac{1}{2}\sum_{i=1}^2\partial_\mu\varphi_i\partial_\nu\varphi_i-\frac{1}{4}Vg_{\mu\nu}-\frac{1}{2}\sum_{i=1}^2X_i^{-2}\left(F_{i\mu\rho}F_{i\nu}\,^\rho-\frac{1}{8}g_{\mu\nu}F_{i\rho\sigma}F_i\,^{\rho\sigma}\right) \notag \\
-&\frac{m^2}{4}\left(X_1X_2\right)^{-1}\left(B_{\mu\rho}B_\nu\,^\rho-\frac{1}{8}g_{\mu\nu}B_{\rho\sigma}B^{\rho\sigma}\right)-\frac{1}{16}\left(X_1X_2\right)^2\left(H_{\mu\rho\sigma}H_\nu\,^{\rho\sigma}-\frac{1}{6}g_{\mu\nu}H_{\rho\sigma\lambda}H^{\rho\sigma\lambda}\right)\,=\,0\,,
\end{align}
\begin{align}
\frac{1}{\sqrt{-g}}\partial_\mu\left(\sqrt{-g}g^{\mu\nu}\partial_\nu\varphi_1\right)\,-&\,\frac{\partial{V}}{\partial\varphi_1}-\frac{1}{2\sqrt{2}}X_1^{-2}F_{1\mu\nu}F_1\,^{\mu\nu}+\frac{1}{2\sqrt{2}}X_2^{-2}F_{2\mu\nu}F_2\,^{\mu\nu}\,=\,0\,, \notag \\
\frac{1}{\sqrt{-g}}\partial_\mu\left(\sqrt{-g}g^{\mu\nu}\partial_\nu\varphi_2\right)\,-&\,\frac{\partial{V}}{\partial\varphi_2}-\frac{1}{4\sqrt{2}}X_1^{-2}F_{1\mu\nu}F_1\,^{\mu\nu}-\frac{1}{4\sqrt{2}}X_2^{-2}F_{2\mu\nu}F_2\,^{\mu\nu} \notag \\
-&\frac{m^2}{8\sqrt{2}}\left(X_1X_2\right)^{-1}B_{\mu\nu}B^{\mu\nu}+\frac{1}{24\sqrt{2}}\left(X_1X_2\right)^2H_{\mu\nu\rho}H^{\mu\nu\rho}\,=\,0\,, 
\end{align}
\begin{align}
\mathcal{D}_\nu\left(X_1^{-2}F_1^{\nu\mu}\right)\,=&\,\frac{1}{24}\sqrt{-g}\varepsilon^{\mu\nu\rho\sigma\tau\lambda}F_{2\nu\rho}H_{\sigma\tau\lambda}\,, \notag \\
\mathcal{D}_\nu\left(X_2^{-2}F_2^{\nu\mu}\right)\,=&\,\frac{1}{24}\sqrt{-g}\varepsilon^{\mu\nu\rho\sigma\tau\lambda}F_{1\nu\rho}H_{\sigma\tau\lambda}\,, \notag \\
\mathcal{D}_\nu\left(\left(X_1X_2\right)^{-1}B^{\nu\mu}\right)\,=&\,\frac{1}{24}\sqrt{-g}\varepsilon^{\mu\nu\rho\sigma\tau\lambda}B_{\nu\rho}H_{\sigma\tau\lambda}\,, \notag \\
\mathcal{D}_\rho\left(\left(X_1X_2\right)^2H^{\rho\nu\mu}\right)\,=&\,-\frac{1}{4}\sqrt{-g}\varepsilon^{\mu\nu\rho\sigma\tau\lambda}\left(\frac{m^2}{2}B_{\rho\sigma}B_{\tau\lambda}+F_{i\rho\sigma}F_{i\tau\lambda}\right)-2m^2\left(X_1X_2\right)^{-1}B^{\mu\nu}\,.
\end{align}

\section{Consistent truncations of \cite{MatthewCheung:2019ehr}} \label{appB}

In this appendix, we briefly review the consistent truncation of seven-dimensional maximal gauged supergravity, \cite{Pernici:1984xx}, on a Riemann surface in \cite{MatthewCheung:2019ehr} and explain the setup to uplift our solutions by employing the truncation ansatz.

The consistent truncation ansatz for the seven-dimensional metric on a Riemann surface, $\Sigma_{\mathfrak{g}}$, is given by
\begin{equation}
ds_7^2\,=\,e^{-4\varphi}ds_5^2+\frac{1}{g^2}e^{6\varphi}ds_{\Sigma_{\mathfrak{g}}}^2\,,
\end{equation}
which introduces a scalar field, $\varphi$, in five dimensions. Also $g^2L_{AdS_5}^2=2^{4/3}$ for the gauge coupling, $g$, and the radius of asymptotic $AdS_5$, $L_{AdS_5}$. The $SO(5)$ gauge fields are decomposed by $SO(5)\rightarrow\,SO(2)\times\,SO(3)$,
\begin{align}
A^{ab}\,=&\,\epsilon^{ab}A+\frac{1}{g}\omega^{ab}\,, \notag \\
A_{a\alpha}\,=&\,-A^{\alpha{a}}\,=\,\psi^{1\alpha}e^a-\psi^{2\alpha}\epsilon^{ab}e^b\,, \notag \\
A^{\alpha\beta}\,=&\,A^{\alpha\beta}\,,
\end{align}
where $a,\,b=1,\,2$, $\alpha,\,\beta=3,\,4,\,5$, $ds_{\Sigma_{\mathfrak{g}}}^2\,=\,e^ae^a$, and $\omega^{ab}$ is the spin connection on $\Sigma_{\mathfrak{g}}$. The ansatz introduces an $SO(2)$ one-form, $A$, $SO(3)$ one-forms, $A^{\alpha\beta}$, transforming in the $(\mathbf{1},\mathbf{3})$ of $SO(2)\times\,SO(3)$, and six scalar fields, $\psi^{a\alpha}=\left(\psi^{1\alpha},\psi^{2\alpha}\right)$, transforming in the $(\mathbf{2},\mathbf{3})$. The scalar fields are given by
\begin{equation}
T^{ab}\,=\,e^{-6\lambda}\delta^{ab}\,, \qquad T^{a\alpha}\,=\,0\,, \qquad T^{\alpha\beta}\,=\,e^{4\lambda}\mathcal{T}^{\alpha\beta}\,, 
\end{equation}
which introduces a scalar field, $\lambda$, and five scalar fields in $\mathcal{T}^{\alpha\beta}$ which live on the coset manifold, $SL(3)/SO(3)$. The three-form field is given by
\begin{align}
S^a\,=&\,K_{(2)}^1\wedge\,e^a-\epsilon^{ab}K_{(2)}^2\wedge\,e^b\,, \notag \\
S^\alpha\,=&\,h_{(3)}^\alpha+\chi_{(1)}^\alpha\wedge\text{vol}_{\Sigma_{\mathfrak{g}}}\,,
\end{align}
which introduces an $SO(2)$ doublet of two-forms, $K_{(2)}^a$, three-forms, $h_{(3)}^\alpha$, and one-forms, $\chi_{(1)}^\alpha$.

To be particular, we consider a subtruncation of the general consistent truncations which reduces to $SU(2)\times\,U(1)$-gauged $\mathcal{N}=4$ supergravity in five dimensions, \cite{Romans:1985ps}, which is presented in section 5.1 of \cite{MatthewCheung:2019ehr}. In this case, we have the scalar fields to be
\begin{equation}
\lambda\,=\,3\varphi\,, \qquad \mathcal{T}_{\alpha\beta}\,=\,\delta_{\alpha\beta}\,, \qquad \psi^{a\alpha}\,=\,0\,.
\end{equation}
From the three-form field, we have a complex two-form field,
\begin{equation}
\mathcal{C}_{(2)}\,=\,K_{(2)}^1+iK_{(2)}^2\,,
\end{equation}
and a three-form field,
\begin{equation} \label{threecon}
*h_{(3)}^\alpha\,=\,\frac{1}{2}e^{-20\varphi}\epsilon_{\alpha\beta\gamma}F^{\beta\gamma}\,,
\end{equation}
with $\chi_{(1)}^\alpha\,=\,0$.

In order to match with the special case of $U(1)^2$-gauged supergravity in seven dimensions, \eqref{specialcase}, we further impose $A_{(1)}^{a\alpha}\,=\,0$ and $\mathcal{C}_{(2)}\,=\,0$. In $U(1)^2$-gauged supergravity in seven dimensions, the scalar fields of are given by
\begin{equation}
T_{ij}\,=\,\text{diag}\left(e^{2\lambda_1}\,,e^{2\lambda_1}\,,e^{2\lambda_2}\,,e^{2\lambda_2}\,,e^{-4\lambda_1-4\lambda_2}\right)\,.
\end{equation}
By matching it with the consistent truncation ansatz,
\begin{equation}
T_{ij}\,=\,\text{diag}\left(e^{-6\lambda}\,,e^{-6\lambda}\,,e^{4\lambda}\,,e^{4\lambda}\,,e^{4\lambda}\right)\,,
\end{equation}
we identify the scalar fields to be
\begin{equation}
\lambda_1\,=\,-3\lambda\,, \qquad \lambda_2\,=\,2\lambda\,.
\end{equation}
The non-trivial three-form field, $S^5$, is given by $h_{(3)}^\alpha$ in \eqref{threecon}.

Finally, we compare the actions of $SU(2)\times\,U(1)$-gauged $\mathcal{N}=4$ supergravity in five dimensions, \cite{Romans:1985ps}, presented in (5.4) of \cite{MatthewCheung:2019ehr} and in (2.1) with (3.1) of \cite{Boido:2021szx} to fix
\begin{equation}
X^{(1)}\,=\,2^{-1/3}e^{10\varphi}\,, \qquad X^{(2)}\,=\,2^{-1/3}e^{10\varphi}\,, \qquad X^{(3)}\,=\,2^{2/3}e^{-20\varphi}\,.
\end{equation}
With $X=X^{(1)}=X^{(2)}$, this determines the scalar fields to be
\begin{equation}
e^{-\frac{10}{9}\lambda_1}\,=\,2^{1/3}X\,, \qquad e^{\frac{5}{3}\lambda_2}\,=\,2^{1/3}X\,, \qquad e^{10\varphi}\,=\,2^{1/3}X\,.
\end{equation}

\bigskip

\bibliographystyle{JHEP}
\bibliography{20220615}

\end{document}